\documentclass[12pt]{article}
\usepackage{amsmath}
\usepackage{amsfonts}
\usepackage{amssymb}
\usepackage{dsfont}
\usepackage[latin1]{inputenc}
\usepackage[all]{xy}
\usepackage{graphicx}
\usepackage{latexsym}
\usepackage{color}
\usepackage{epsfig}
\usepackage{graphicx}
\usepackage{caption}
\usepackage{subcaption}
\usepackage{here}
\usepackage{float}
\usepackage{array}
\usepackage{tikz}
\usepackage{tikz-cd}
\usetikzlibrary{3d,arrows,calc,chains,matrix,positioning,scopes}

\def\l[{\left[}
\def\r]{\right]}

\def\ba{\begin{array}}
\def\ea{\end{array}}
\def\beq{\begin{equation}}
\def\eeq{\end{equation}}
\def\bea{\begin{eqnarray}}
\def\eea{\end{eqnarray}}

\newcommand{\Hom}{\mathop{}\mathopen{}{\rm Hom}\!}
\newcommand{\End}{\mathop{}\mathopen{}{\rm End}\!}

\newcommand{\Ext}{\mathop{}\mathopen{}{\rm Ext}}

\begin{document}

\pagestyle{empty}
\setcounter{page}{0}
\hspace{-1cm}

\begin{center}
{\Large {\bf Abelian Turaev-Virelizier theorem and $U(1)$ BF surgery formulas}}%
\\[1.5cm]

{\large Ph. Mathieu and F. Thuillier}

\end{center}

\vskip 0.7 truecm

{\it LAPTH, Université Savoie Mont Blanc, CNRS, 9, Chemin de Bellevue, BP 110, F-74941
Annecy-le-Vieux cedex, France.}

\vspace{3cm}

\centerline{{\bf Abstract}}
In this article we construct the Reshetikhin-Turaev invariant associated with the Drinfeld Center of the spherical category arising from the $U(1)$ BF theory on a closed $3$-manifold $M$. This invariant is shown to coincide with the Turaev-Viro invariant of $M$ thus providing an example of the Turaev-Virelizier theorem. Finally we exhibit some surgery formulas for the abelian Turaev-Viro invariant which are very similar to the surgery formulas of the abelian Reshetikhin-Turaev invariant obtained in the $U(1)$ Chern-Simons context.

\vspace{2cm}


\newpage
\pagestyle{plain} \renewcommand{\thefootnote}{\arabic{footnote}}

\section{Introduction}
The relation between Quantum Field Theory and $3$-manifold invariants highlighted by E. Witten \cite{W1} has been meticulously investigated in the $U(1)$ case for Chern-Simons and BF theories thanks to the use of Deligne-Beilinson cohomology \cite{GT1,GT2,GT3,MT1,MT2}. In particular it was proven that the partition functions of these Quantum Field Theories coincide with the Reshetikhin-Turaev \cite{RT} and Turaev-Viro \cite{TV} invariants respectively of $M$. The categories on which these invariants are built from are nothing but the irreducible representations of $\mathbb{Z}_k$, $k$ being the coupling constant of the $U(1)$ CS and BF theories. So the cyclic group $\mathbb{Z}_k$ can be seen as the abelian equivalent of the Quantum Group appearing in the $SU(2)$ case.

As a modular category the Drinfeld Center of the spherical category used to generate the Turaev-Viro invariant can be used to provide a Reshetikhin-Turaev invariant. The conjecture that these two invariants coincide has been recently turned into a proved theorem by V. Turaev and A. Virelizier \cite{TuVi}.

In this article we will show how the $U(1)$ BF partition function is related with the Reshetikhin-Turaev invariant of the Drinfeld center of $\mathbb{Z}_k$, thus getting an explicit example of Turaev-Virelizier theorem as well as, by combining this result with the one of \cite{MT2}, a reciprocity formula.
Eventually we show how to obtain a surgery formula for the Turaev-Viro invariant from the BF point of view, that is to say how to determine this invariant for a closed $3$-manifold $M$ from a computation in $S^3$ once a surgery link of $M$ in $S^3$ as been provided. We will show that:
\bea
\Upsilon_k(M) = \frac{1}{k^m} \langle \widehat{W}_{S^3}({\mathcal{L}_M};\overline{\alpha},\overline{\beta}) \rangle_{BF_k} \, ,
\eea
where $\Upsilon_k(M)$ is the ``specially normalized" abelian Turaev-Viro invariant of $M$ \cite{MT1} and $\widehat{W}_{S^3}({\mathcal{L}_M};\overline{\alpha},\overline{\beta})$ is the $U(1)$ BF surgery function of $M$. Ultimately a surgery invariant for any link $L$ in $M$ as well as a surgery formula associated with this surgery invariant will be exhibited.

\vspace{2mm}
All along this article $M$ denotes a closed oriented $3$-manifold and $\Pi_M$ is a \emph{good} cellular decomposition of $M$, that is to say a cellular decomposition such that $H_k(\Pi_M) \cong H_k(M)$ for all $k$.

\section{$U(1)$ BF partition function and Turaev-Viro invariant}

While choosing $\int A \wedge dB$ as $U(1)$ BF action on $\mathds{R}^3$ is legitimate, this expression does not extend to closed oriented $3$-manifolds. Indeed on such manifolds $U(1)$-connections are not globally defined $1$-forms. Furthermore the group of gauge transformations which is $\Omega^0(M)/\mathds{R}$ on $\mathds{R}^3$ becomes $\Omega^1_{\mathds{Z}}(M)$, the space of closed $1$-forms with integral periods on $M$. In order to get ride of gauge invariance it seems natural to try to deal with gauge classes of $U(1)$-connections instead of local representatives. It so happens that the first Deligne-Beilinson cohomology group, $H_D^1(M)$, canonically identifies with the set of equivalence classes of $U(1)$-bundles with connections on $M$. Hence $H_D^1(M)$ is the appropriate set of ``fields" to consider in the $U(1)$ BF theory. It has to be noted that $H_D^1(M)$ is also the appropriate set of fields for the $U(1)$ Chern-Simons theory \cite{GT1}.

The group $H_D^1(M)$ is embedded into the following exact sequence:
\bea
\label{sequence1}
0 \longrightarrow {\Omega^1( M) \over {\Omega_{\mathbb{Z}}^1(M)}} \longrightarrow H_D^1(M,\mathbb{Z})
\longrightarrow H^{2}(M,\mathbb{Z}) \longrightarrow 0 \, ,
\eea

\noindent This exact sequence allows to decompose (non canonically) each $A \in H_D^1(M)$ according to $A = A_{\vec{n}} + \overline{\alpha}$, where $A_{\vec{n}}$ is an origin on the fiber of $H_D^1(M)$ over $\vec{n} \in H^{2}(M,\mathbb{Z})$ and $\overline{\alpha} \in  {\Omega^1( M) \over {\Omega_{\mathbb{Z}}^1(M)}}$. In fact as $H^{2}(M,\mathbb{Z}) = F^2(M) \oplus T^2(M)$ we can even write $A_{\vec{n}} = A_{\vec{u}} + A_{\vec{\tau}}$ with $\vec{u} \in F^{2}(M)$ and $\vec{\tau} \in T^{2}(M)$. The set of fields of the $U(1)$ BF theory on $M$ is then chosen to be $H_D^1(M)$\footnote{It can alternatively be chosen as the Pontrjagin dual of $H_D^1(M)$ which has a similar structure except that its fibers are made of distributional classes instead of smooth ones. This is quite irrelevant for our purpose here and we send the reader to \cite{GT1,MT1} for details.}.

The group $H_D^1(M)$ can be endowed with a commutative product, $\star$, for which:
\bea
\label{action}
\int_M A \star B \in \mathds{R}/\mathds{Z} \, ,
\eea

\noindent for any $A,B \in H_D^1(M)$. Naively, the product $A \star B$ can be written as $\theta \eta_M$ with $\theta \in  \mathds{R}/\mathds{Z}$ and $\eta_M$ a normalized volume form on $M$, and locally, i.e. in any contractible open subset of $M$, $A \star B = \omega_A \wedge d \omega_B$ for some local $1$-forms $\omega_A$ and $\omega_B$. The $U(1)$ BF action with coupling constant $k$ is then:
\bea
\label{k-action}
k \int_M A \star B \in \mathds{R}/\mathds{Z} \, ,
\eea

\noindent and due to (\ref{action}) we deduce that $k \in \mathds{Z}$ (i.e. the coupling constant is quantized as in the $U(1)$ Chern-Simons theory \cite{GT1}).

The partition function of the Quantum Field Theory defined by the $U(1)$ BF action with coupling constant $k$ formally reads:
\bea
\label{bfpartition}
Z_{BF_{k}}(M) \equiv {\sum_{\vec{m},\vec{n} \in H^2(M)}} { \int \int D\overline{\alpha} \;  D\overline{\beta} \cdot \exp\left\{{2i\pi k \int_M \left( (A_{\vec{m}} + \overline{\alpha}) \star (A_{\vec{n}} + \overline{\beta}) \right) } \right\} \over \int \int D\overline{\alpha} \; D\overline{\beta} \cdot \exp\left\{{2i\pi k \int_M \overline{\alpha} \star \overline{\beta}  } \right\} } \, .
\eea

\noindent The computation fully detailed in \cite{MT1} yields:
\begin{equation}
\label{bfpartition2}
Z_{BF_{k}}(M)  =  \sum_{\vec{\kappa} \in T^2} \sum_{\vec{\tau} \in T_1}  e^{- 2 \pi i k Q(\vec{\kappa},\vec{\tau}) } \, ,
\end{equation}

\noindent where $Q: T^2(M) \times T^2(M) \rightarrow \mathds{R}/\mathds{Z}$ is the linking form, which after computation gives:
\begin{equation}
\label{bfpartition3}
Z_{BF_{k}} = \prod\limits_{j=1}^{N}\gcd\left(k,p_{j}\right)p_{j} \, ,
\end{equation}
where $T^2(M)$ has been decomposed according to $T^2(M) \cong \mathds{Z}_{p_1} \oplus \cdots \oplus \mathds{Z}_{p_N}$, with $p_j | p_{j+1}$.

\vspace{2mm}

It was shown in \cite{MT1} that the set of representations of $\mathds{Z}_k$ plays the role of the spherical category on which a Turaev-Viro construction can be applied. More explicitly the objects of the category $\mathbb{C}^{\mathbb{Z}_k}$ under consideration are the irreducible representations of $\mathbb{Z}_k$. We denote these objects by $R_p$ with $p=0, \cdots, k-1$. The unit object is the trivial representation $R_0$ and the unit morphism, denoted by $I\!d_p$, is just multiplication by $1$.
As usual, natural transformations are the morphisms of this category and hence:
\bea
\label{decadix}
\Hom \, (R_p,R_q) = \delta_{p,q} \End \, \mathbb{C} \cong \delta_{p,q} \, \mathbb{C} = \delta_{p-q,0} \, \mathbb{C} \, .
\eea

The category $\mathbb{C}^{\mathbb{Z}_k}$ is trivially turned into a tensor category by noticing that $R_p \otimes R_q \cong R_{p+q} \cong R_q \otimes R_p$. Duality is also trivially defined by $(R_p)^* = R_{-p} = R_{k-p}$, since we work modulo $k$. So $\mathbb{C}^{\mathbb{Z}_k}$ is a \emph{pivotal} category.

The left and right traces are also trivial in $\mathbb{C}^{\mathbb{Z}_k}$ and they coincide so that finally $\mathbb{C}^{\mathbb{Z}_k}$ is a \emph{spherical} category.

Once a good cellular decomposition $\Pi_M$ of $M$ is provided we can apply one of the standard constructions \cite{TV,BW,BK} to generate a Turaev-Viro invariant of $M$. First we introduce the notion of $\mathds{Z}_k$-\emph{labeling} of the edges of $\Pi_M$ that is to say an assignment of an element of $\mathds{Z}_k$ to each edge of $\Pi_M$. Since any face (i.e. $2$-cell) of $\Pi_M$ is bounded by edges of $\Pi_M$, any $\mathds{Z}_k$-labeling of $\Pi_M$ canonically defines a $\mathds{Z}_k$-labeling of the faces of $\Pi_M$. More precisely, given a labeling $l$ of the edges of $\Pi_M$, we associate to any oriented face $F$ bounded by the $n_F$ edges $\sigma_i$ the $\mathds{Z}_k$-valued quantity $\Sigma^l_{F} = \sum\limits_{i=1}^{n_{F}} l(\sigma_i)= \sum\limits_{i=1}^{n_{F}} l_i$. The \emph{state spaces} of the construction are then:
\bea
\label{statespace}
H(F,l_\Pi) = \Hom\,(R_0,R_{l_1} \otimes \cdots \otimes R_{l_{n_{F}}}) = \delta_{ \Sigma^l_{F} ,0 } \, \mathbb{C} \, ,
\eea
and the $\mathds{Z}_k$ Turaev-Viro invariant of $M$ is defined as:
\bea
\label{TVab}
\Upsilon_{k}(M) = k^{-(v-1)} \sum_l \left( \prod_F \delta_{ \Sigma^l_{F} , 0} \right) \, ,
\eea

\noindent Strictly speaking the normalization factor usually taken is $k^{-v}$ rather than $k^{-(v-1)}$ so that $\Upsilon_{k}(M)$ is related to the standard Turaev-Viro invariant $\tau_{k}(M)$ according to $\Upsilon_{k}(M) = k. \tau_{k}(M)$.

On the first\emph{} hand a simple computation yields:
\bea
\label{TVfinalresult}
\Upsilon_{k}(M) = |H^1(M,\mathbb{Z}_k)| \, ,
\eea

\noindent and on the second hand it is easy to check that: $|H^1(M,\mathbb{Z}_k)| = k^{b_1} \prod\limits_{j=1}^{N}\gcd\left(k,p_{j}\right)$ where $b_1$ is the first Betti number of $M$.

\noindent We finally get:
\bea
\label{TvvsBF}
\Upsilon_k(M) = \frac{k^{b_1}}{p_1 \ldots p_d} Z_{BF_{k}}   \, .
\eea

Let us go backtrack on the difference of normalization between the invariantd $\Upsilon_k(M)$ and $\tau_k(M)$. In the abelian context the choice made in \cite{MT1} which leads to $\Upsilon_k(M)$ seems more natural since with convention (\ref{TVab}) we get relation (\ref{TVfinalresult}) whereas the Turaev-Viro convention yields $\tau_{k}(M) = \frac{1}{k} |H^1(M,\mathbb{Z}_k)|$. In particular we find that $\tau_{k}(S^1 \times S^2) = 1$ and $\tau_{k}(S^3) = 1/k$, whereas $\Upsilon_k(S^1 \times S^2) = k$ and $\Upsilon_k(S^3) = 1$. Besides in definition (\ref{bfpartition}) of the $U(1)$ BF partition function the normalization factor deals with the fiber over $0 \in H^2(M)$ which turns out to be the unique fiber of $H^1_D(S^3)$. To that extend the normalization for the $U(1)$ BF partition function is taken with respect to $S^3$ and hence in both sides of relation (\ref{TvvsBF}) $S^3$ is taken as reference manifold. The same difference in normalization occurs when trying to relate the $U(1)$ Chern-Simons partition function with the $\mathbb{Z}_k$ Reshetikhin-Turaev invariant. In fact this discrepancy in normalization also appears in the non-abelian context: the ``mathematical" normalization is taken with respect to $S^1 \times S^2$ whereas the one coming from Quantum Field Theory is taken with respect to $S^3$. For instance the $SU(2)$ Reshetikhin-Turaev invariant of $S^1 \times S^2$ is $1$ whereas the expectation value -- with respect to the Chern-Simons functional measure -- of the unknot in $S^3$ which is supposed to produce this invariant (perturbatively) yields $q^{-\frac{1}{2}} + q^{\frac{1}{2}}$ \cite{GMM}.

\section{Drinfeld center of $\mathbb{C}^{\mathbb{Z}_k}$}

The Drinfeld center of the spherical category $\mathbb{C}^{\mathbb{Z}_k}$, denoted $\mathcal{Z}(\mathbb{C}^{\mathbb{Z}_k})$, is a category whose objects are couples $(R_p,\sigma)$ where $R_p$ is an object of $\mathbb{C}^{\mathbb{Z}_k}$ and $\sigma$ is a collection of (natural) isomorphisms $\sigma_q : R_p \otimes R_q \rightarrow R_q \otimes R_p$ such that:
\bea
\label{natiso}
\sigma_{q \otimes r} = \sigma_{q + r} = (Id_q \otimes \sigma_{r}) \circ (\sigma_{q} \otimes Id_r) = \sigma_{q} \sigma_{r} \, .
\eea

By taking into account the cyclic character of the construction, we immediately deduce from relation (\ref{natiso}) that:
\bea
\sigma_{q} = e^{2i\pi \frac{qu}{k}} := \sigma_{q}^{(u)} \, ,
\eea

\noindent for some $u \in \mathds{Z}_k$. From now on we denote by $(R_p,\sigma^{(u)})$ an object of $\mathcal{Z}(\mathbb{C}^{\mathbb{Z}_k})$. The collection of these objects is ${\mathbb{Z}_k} \times {\mathbb{Z}_k}$, and by construction $\sigma_{q}^{(u)}$ is a braiding for $\mathbb{C}^{\mathbb{Z}_k}$.

A morphism $f: (R_p,\sigma^{(u)}) \rightarrow (R_q,\sigma^{(v)}) $ of $\mathcal{Z}(\mathbb{C}^{\mathbb{Z}_k})$ is given by an element $f \in \Hom \, (R_p,R_q)$ such that:
\bea
\label{Zmorphism}
(Id_r \otimes f) \circ \sigma_r^{(u)} = \sigma_r^{(v)} \circ (Id_r \otimes f) \, ,
\eea

\noindent for all $r$. In our abelian context this simply gives :
\bea
\label{Zmorphism2}
f \delta_{p,q} e^{2i\pi \frac{ru}{k}} = f \delta_{p,q} e^{2i\pi \frac{rv}{k}} \, ,
\eea
for all $r$, which implies that:
\bea
\label{Zmorphism3}
\Hom ((R_p,\sigma^{(u})),(R_q,\sigma^{(v)})) \cong \delta_{p,q} \delta_{u,v} \mathds{C} \, .
\eea
The Drinfeld center turns into a monoidal category once we have set:
\bea
\label{Zmono}
(R_p,\sigma^{(u)}\emph{}) \otimes (R_q,\sigma^{(v)}) := (R_{p+q},\sigma^{(u+v)}) \, ,
\eea
and duality is simply defined by:
\bea
\label{Zdual}
(R_p,\sigma^{(u)})^* := (R_{k-p},\sigma^{(k-u)}) \, .
\eea

There is a natural braiding on $\mathcal{Z}(\mathbb{C}^{\mathbb{Z}_k})$ given by:
\bea
\label{Zbraid}
C_{(p,u),(q,v)} := \sigma_p^{(v)} = e^{2i\pi \frac{pv}{k}} \, .
\eea

\noindent This braiding is not symmetric since $C_{(q,v),(p,u)} = e^{2i\pi \frac{qu}{k}}$ although it obviously satisfies the usual braiding constraints:
\bea
\label{Zbraidconst}
 \left\{ \begin{gathered}
C_{(p+q,u+v),(r,w)} =  e^{2i\pi \frac{(p+q)w}{k}} = C_{(p,u),(r,w)}C_{(q,v),(r,w)} \\
C_{(p,u),(q+r,v+w)} =  e^{2i\pi \frac{p(v+w)}{k}} = C_{(p,u),(q,v)}C_{(p,u),(r,w)}
\end{gathered} \right. \, ,
\eea
the Yang-Baxter constraint being trivially fulfilled in this abelian context. Similarly there is a natural twist on $\mathcal{Z}(\mathbb{C}^{\mathbb{Z}_k})$ given by:
\bea
\label{Ztwist}
\Theta_{(p,u)} := \sigma_p^{(u)} = e^{2i\pi \frac{pu}{k}} \, .
\eea
This morphism is actually a twist as it satisfies:
\bea
\label{Ztwist}
\Theta_{(p+q,u+v)} = e^{2i\pi \frac{(p+q)(u+v)}{k}} =  C_{(q,v),(p,u)} C_{(p,u),(q,v)} \Theta_{(p,u)}\Theta_{(q,v)} \, ,
\eea
and it is compatible with duality since:
\bea
\label{Ztwistdual}
\Theta_{(p,u)^*} = \Theta_{(k-p,k-u)} = e^{2i\pi \frac{(k-p)(k-u)}{k}} = \Theta_{(p,u)} \, .
\eea

\noindent The braiding on $\mathcal{Z}(\mathbb{C}^{\mathbb{Z}_k})$ is also compatible with duality since it fulfills:
\bea
\label{Zribbon}
C_{(p,u),(q,v)^*} = e^{2i\pi \frac{p(k-v)}{k}} = (C_{(p,u),(q,v)})^{-1}  \, .
\eea
all these properties provide $\mathcal{Z}(\mathbb{C}^{\mathbb{Z}_k})$ with the structure of a \emph{Ribbon} category \cite{Tu}. The final step is to show that this Ribbon category is also a \emph{modular} category in the sense of Turaev \cite{Tu}. This is achieved by introducing the $S$-matrix:
\bea
\label{ZSmatrix}
S_{(p,u),(q,v)} := C_{(q,v),(p,u)} C_{(p,u),(q,v)} = e^{2i\pi \frac{qu+pv}{k}} \, .
\eea

\noindent This matrix is symmetric and after some columns and rows rearrangement we have:
\bea
\label{detZSmatrix}
det(S) = \pm det(A) \, ,
\eea

\noindent where:
\bea
\label{A}
A = \left( {\begin{array}{*{20}{c}}
  1&1&1& \cdots &1&1 \\
  1&\alpha &{{\alpha ^2}}& \cdots &{{\alpha ^{k - 2}}}&{{\alpha ^{k - 1}}} \\
  1&{{\alpha ^2}}&{{\alpha ^4}}& \cdots &{{\alpha ^{(k - 2)2}}}&{{\alpha ^{2(k - 1)}}} \\
   \vdots & \vdots & \vdots & \ddots & \vdots & \vdots  \\
  1&{{\alpha ^{k - 2}}}&{{\alpha ^{2(k - 2)}}}& \cdots &{{\alpha ^{(k - 2)\left( {k - 2} \right)}}}&{{\alpha ^{(k - 1)\left( {k - 2} \right)}}} \\
  1&{{\alpha ^{k - 1}}}&{{\alpha ^{2(k - 1)}}}& \cdots &{{\alpha ^{(k - 2)\left( {k - 1} \right)}}}&{{\alpha ^{(k - 1)\left( {k - 1} \right)}}}
\end{array}} \right)\ \, ,
\eea
where $\alpha  = {e^{\frac{{2i\pi }}{k}}}$. This is nothing but a Vandermonde matrix and hence:
\bea
\label{ZSok}
det(S) = \pm {\left( {\prod\limits_{0 \leqslant m < n \leqslant k - 1} {\left( {{e^{2i\pi \frac{n}{k}}} - {e^{2i\pi \frac{m}{k}}}} \right)} } \right)^{2k}} \ne 0\ \, .
\eea
\noindent Let us point out that  the determinant of the $S$-matrix of $\mathbb{C}^{\mathbb{Z}_k}$ - the Ribbon category which gives rise to the abelian Reshetikhin-Turaev invariant of $M$ -- is also given by a Vandermonde determinant \cite{MT1}. This determinant vanishes if and only if $2(m-n) = 0 \; [k]$, an equation which has non trivial solutions when $k$ is even. Accordingly the Ribbon category $\mathbb{C}^{\mathbb{Z}_k}$ is modular if and only if $k$ is odd, whereas $\mathcal{Z}(\mathbb{C}^{\mathbb{Z}_k})$ is a Ribbon category for any $k$.

The dimension of an object $(R_p,\sigma^{(u)})$ of the modular category $\mathcal{Z}(\mathbb{C}^{\mathbb{Z}_k})$ is then given by:
\bea
\label{dim}
\dim((R_p,\sigma^{(u)})) = S_{(p,u),(0,0)} = e^{2i\pi \frac{0u+p0}{k}} = 1 \, ,
\eea
and the dimension of $\mathcal{Z}(\mathbb{C}^{\mathbb{Z}_k})$ by:
\bea
\label{dimZ}
\mathcal{D} = dim(\mathcal{Z}(\mathbb{C}^{\mathbb{Z}_k})) := \sqrt {{{\sum\limits_{p,u = 0}^{k - 1} ( \dim \left( {{R_p},{\sigma ^{\left( u \right)}}} \right) )^2 } }} = \sqrt {{k^2}}  = k \, .
\eea

\section{$\mathcal{Z}(\mathbb{C}^{\mathbb{Z}_k})$ Reshetikhin-Turaev invariant of $M$ and surgery formula}

\noindent The normalisation factor $\Delta_k$ appearing in the Reshetikhin-Turaev construction \cite{RT,Tu} (see also \cite{MOO}) is:
\bea
\label{dimZ}
\Delta_k = \sum\limits_{p,u = 0}^{k - 1} {{e^{2i\pi \frac{{pu}}{k}}}} = k = \mathcal{D} \, .
\eea

The Reshetikhin-Turaev invariant of $M$ generated by the Ribbon category $\mathcal{Z}(\mathbb{C}^{\mathbb{Z}_k})$ is then:
\bea
\label{ZRT}
R{T_{Z\left( {{\mathbb{C}^{{\mathbb{Z}_k}}}} \right)}}\left( M \right) = \Delta _k^{s\left( L \right)}{D^{ - s\left( L \right) - m - 1}}\sum\limits_{\left( {p,u} \right) \in {{\left( {{\mathbb{Z}_k} \times {\mathbb{Z}_k}} \right)}^m}} {{e^{2i\pi \frac{{{p^i}{L_{ij}}{u^j}}}{k}}}} \, ,
\eea

\noindent where $(L_{ij})$ is the $m \times m$ linking matrix of a surgery link $\mathcal{L}$ of $M$ in $S^3$. Even if its expression relies on $\mathcal{L}$, the invariant $R{T_{Z\left( {{\mathbb{C}^{{\mathbb{Z}_k}}}} \right)}}\left( M \right)$ depends on $M$ and not on the surgery link representing $M$.

Taking into account the previous relations we finally obtain:
\bea
\label{ZRT}
R{T_{Z\left( {{\mathbb{C}^{{\mathbb{Z}_k}}}} \right)}}\left( M \right) = \frac{1}{{{k^{m + 1}}}}\sum\limits_{\left( {p,u} \right) \in {{\left( {{\mathbb{Z}_k} \times {\mathbb{Z}_k}} \right)}^m}} {{e^{2i\pi \frac{{{p^i}{L_{ij}}{u^j}}}{k}}}} \, .
\eea
A straightforward computation leads to:
\bea
\label{Zresult1}
R{T_{Z\left( {{\mathbb{C}^{{\mathbb{Z}_k}}}} \right)}}\left( M \right) = \frac{1}{k}\sum\limits_{\vec u \in {{\left( {{\mathbb{Z}_k}} \right)}^m}} {\delta _{L\vec u,\vec 0}^{\left[ k \right]}} \, ,
\eea

\noindent that is to say:
\bea
\label{Zresult2}
R{T_{Z\left( {{\mathbb{C}^{{\mathbb{Z}_N}}}} \right)}}\left( M \right) = \frac{1}{k}\left| {\ker {L^{\left[ k \right]}}} \right| \, ,
\eea

\noindent where $L^{\left[ k \right]} : (\mathds{Z}_k)^m \rightarrow (\mathds{Z}_k)^m$ denotes the linear morphism canonically associated with the linear morphism $L : \mathds{Z}^m \rightarrow \mathds{Z}^m$ of the linking matrix of $\mathcal{L}$. There is a well-known exact sequence for $L$:
\bea
\label{Lexactseq}
0 \to \ker L \to {\mathbb{Z}^m}\mathop  \to \limits^L {\mathbb{Z}^m} \to {\text{Coker}}\;L \to 0 \, ,
\eea
which induces a dual exact sequence (standard property of the $\Hom$ functor) which on its turn yields:
\bea
0 \to \ker {L^{\left[ k \right]}} \to {\left( {{\mathbb{Z}_k}} \right)^m}\mathop  \to \limits^{{L^{\left[ k \right]}}} {\left( {{\mathbb{Z}_k}} \right)^m} \to {\text{Coker}}\;{L^{\left[ k \right]}} \to 0 \, ,
\eea
with $\ker {L^{\left[ k \right]}} \cong \Hom\left( {{\text{Coker}}\;L,{\mathbb{Z}_k}} \right)$. Since we have:
\bea
{\text{Coker}}\;L \cong {H_1}\left( M \right) \, ,
\eea
and:
\bea
{H^1}\left( {M,{\mathbb{Z}_k}} \right) \cong \Hom\left( {{H_1}\left( M \right),{\mathbb{Z}_k}} \right) \oplus \Ext\left( {{H_0}\left( M \right),{\mathbb{Z}_k}} \right) = \Hom\left( {{H_1}\left( M \right),{\mathbb{Z}_k}} \right) \, ,
\eea
we deduce that:
\bea
\label{ZResult2}
R{T_{Z\left( {{\mathbb{C}^{{\mathbb{Z}_k}}}} \right)}}\left( M \right) = \frac{1}{k} \left | {{H^1}\left( {M,{\mathbb{Z}_k}} \right)} \right| \, .
\eea

Comparing this result with the one obtained in \cite{MT1} we conclude that:
\bea
\label{Tureav-Virelilzier}
R{T_{Z\left( {{\mathbb{C}^{{\mathbb{Z}_k}}}} \right)}}\left( M \right) = \frac{1}{k} \Upsilon_k(M) = \tau_k(M) \, ,
\eea
which yields the abelian Turaev-Virelizier theorem.

Let us combine all this with results obtained in \cite{MT2} and write reciprocity formulas thus generated:
\bea
\label{final}
\begin{aligned}
\frac{1}{{{k^m}}}\sum\limits_{\vec p \in {{\left( {{\mathbb{Z}_k}} \right)}^m}} {\;\sum\limits_{\vec u \in {{\left( {{\mathbb{Z}_k}} \right)}^m}} {{e^{2i\pi \frac{{{p^i}{L_{ij}}{u^j}}}{k}}}} }  =  & \frac{1}{{{k^{F + V - 1}}}}\sum\limits_{\vec q \in {{\left( {{\mathbb{Z}_k}} \right)}^F}} {\;\sum\limits_{\vec v \in {{\left( {{\mathbb{Z}_k}} \right)}^E}} {{e^{2i\pi \frac{{{q^a}{D_{ab}}{v^b}}}{k}}}} }  \\
=  & \frac{{{k^{{b_1}}}}}{{{p_1} \times  \cdots  \times {p_T}}}\sum\limits_{\vec \kappa  \in {T^2}\left( M \right)} {\;\sum\limits_{\vec \tau  \in {T^2}\left( M \right)} {{e^{2i\pi k {\kappa ^\alpha }{Q_{\alpha \beta }}{\tau ^\beta }}}} }  \, , \\
\end{aligned}
\eea

\noindent where $F$ (resp. $V$) is the number of faces (resp. vertices) of the cellular decomposition $\Pi_M$ of $M$, $\left( {{D_{ab}}} \right)$ is the matrix representing the de Rham differential on $1$-cocycles of $\Pi_M$ and $\left( {{Q_{\alpha \beta }}} \right)$ the matrix representing the linking form $Q: T^2(M) \times T^2(M) \rightarrow \mathbb{Q}/\mathbb{Z}$.

In analogy with the abelian Reshetikhin-Turaev surgery formula \cite{GT3} we can wonder whether the left-hand side of equations (\ref{final}) can be seen as coming from a computation in $S^3$. To see this let us consider an integer Dehn framed surgery link $\mathcal{L}_M$ of $M$ in $S^3$ such that $\mathcal{L}_M = \mathcal{L}_1 \cup \mathcal{L}_2 \cup \cdots \mathcal{L}_m$. As $H^2(S^3) = 0$, any class $A \in H^1_D(S^3)$ can be canonically identified with a class $\overline{\alpha} \in \Omega^1(S^3) / \Omega_{\mathbb{Z}}^1(S^3)$ in $S^3$ according to $A = 0 + \overline{\alpha}$. With the introduction of an orientation for each component $\mathcal{L}_i$ of $\mathcal{L}_M$, the $U(1)$ \emph{BF surgery function} $\widehat{W}_{S^3}({\mathcal{L}_M};\overline{\alpha},\overline{\beta})$ of $\mathcal{L}_M$ in $S^3$ is defined as:
\bea
\label{BFsurgfunc}
\widehat{W}_{S^3}({\mathcal{L}_M};\overline{\alpha},\overline{\beta}) = \prod\limits_{j = 1}^{m} \left( \sum\limits_{ p_j = 0}^{k-1} e^{2i\pi p_j \oint_{\mathcal{L}_j} \overline{\alpha} } \sum\limits_{ u_j = 0}^{k-1} e^{2i\pi u_j  \oint_{\mathcal{L}_j} \overline{\beta}} \right) \, ,
\eea
where $e^{2i\pi p_j \oint_{\mathcal{L}_j} \overline{\alpha} }$ and $e^{2i\pi u_j  \oint_{\mathcal{L}_j} \overline{\beta}}$ are the holonomies of the gauge classes $\overline{\alpha}$ and $\overline{\beta}$ along the component loops $\mathcal{L}_j$ with charge $p_j$ and $u_j$ respectively. The expectation value of the $U(1)$ BF surgery function of $\mathcal{L}_M$ in $S^3$ is then:
\bea
\label{surg}
\langle \widehat{W}_{S^3}({\mathcal{L}_M};\overline{\alpha},\overline{\beta}) \rangle_{BF_k} = {\int \int D\overline{\alpha} \; D\overline{\beta} \cdot e^{{2i\pi k \int_M \overline{\alpha} \star \overline{\beta}  } } \; \widehat{W}_{\mathcal{L}_M}(\overline{\alpha},\overline{\beta}) \over \int \int D\overline{\alpha} \; D\overline{\beta} \cdot e^{{2i\pi k \int_M \overline{\alpha} \star \overline{\beta}  } } } \, ,
\eea
the evaluation of which \cite{MT2} yields:
\bea
\label{expSurgfunc}
\langle \widehat{W}_{S^3}({\mathcal{L}_M};\overline{\alpha},\overline{\beta}) \rangle_{BF_k} = \sum\limits_{\vec p \in {{\left( {{\mathbb{Z}_k}} \right)}^m}} \; \sum\limits_{\vec u \in {{\left( {{\mathbb{Z}_k}} \right)}^m}} e^{- \frac{2i\pi}{k} \langle \vec p , \mathbb{L} \vec u \rangle}  \, ,
\eea
where $\mathbb{L}$ is the linking matrix of the surgery link $\mathcal{L}_M$ in $S^3$. The minus sign in the exponential is obviously irrelevant so that putting (\ref{expSurgfunc}) all together with (\ref{final}) we get:
\bea
\label{surgTV}
\Upsilon_k(M) = \frac{1}{k^m} \langle \widehat{W}_{S^3}({\mathcal{L}_M};\overline{\alpha},\overline{\beta}) \rangle_{BF_k} \, ,
\eea
which provides a surgery formula for the abelian Turaev-Viro invariant analogous to the abelian Reshetikhin-Turaev surgery formula obtained from the $U(1)$ Chern-Simons theory \cite{GT3}. By comparing relations (\ref{surgTV}) and (\ref{TvvsBF}) we can notice that unlike the former the latter requires a normalization factor which depends on $M$ in order to provide the abelian Turaev-Viro invariant of $M$. The normalization factor $\frac{1}{k^m}$ in the right-hand side of the BF surgery formula (\ref{surgTV}) simply ensures that the resulting expression depends on $M$ and not on the integer Dehn framed surgery link of $\mathcal{L}_M$. Since the same phenomena holds in the $U(1)$ Chern-Simon case \cite{GT3} we can say that in the context of these $U(1)$ Quantum Field Theory the surgery formula is a little more efficient than the direct computation on $M$.

As an example let us consider the unknot with zero charge in $S^3$ is a surgery link for $S^1 \times S^2$. A simple computation shows that $\widehat{W}_{S^3}({\mathcal{L}_{S^1 \times S^2}};\overline{\alpha},\overline{\beta}) \rangle_{BF_k} = k^2$ and hence that $\Upsilon_k(S^1 \times S^2) = k$ which is the correct result. More generally the unknot with charge $p$ in $S^3$ provides a surgery link for the lens space $L(p,1)$. We have $\widehat{W}_{S^3}({\mathcal{L}_{L(p,1)}};\overline{\alpha},\overline{\beta}) \rangle_{BF_k} = k \gcd(k,p)$ and thus $\Upsilon_k(L(p,1)) = \gcd(k,p)$ which is the right answer \cite{MT1}.

Finally if $L$ denotes a link in the complement of $\mathcal{L}_M$ in $S^3$, then it defines a link in $M$, still denoted $L$, and we have the more general surgery formula:
\bea
\label{surgInv}
\langle W_M(L;A,B) \rangle_{BF_k} = { \langle \widehat{W}_{S^3}({\mathcal{L}_M};\overline{\alpha},\overline{\beta}) \, W_{S^3}(L;\overline{\alpha},\overline{\beta}) \rangle_{BF_k}  \over \langle \widehat{W}_{S^3}({\mathcal{L}_M};\overline{\alpha},\overline{\beta}) \rangle_{BF_k} } \, ,
\eea
where $\langle W_M(L;\overline{\alpha},\overline{\beta}) \rangle_{BF_k}$ is the BF expectation value of the holonomies $W_M(L;\overline{\alpha},\overline{\beta}) = e^{2i\pi \oint_L \overline{\alpha} } e^{2i\pi \oint_L \overline{\beta} }$ in $S^3$ whereas $\langle W_M(L;A,B) \rangle_{BF_k}$ is the BF expectation value of the holonomies $W_M(L;A,B) = e^{2i\pi \oint_L A } e^{2i\pi \oint_L B}$ in $M$. In particular, the quantity:
\bea
\label{surginvL}
\Upsilon_k(M;L) = \frac{1}{k^m} \, \langle \widehat{W}_{S^3}({\mathcal{L}_M};\overline{\alpha},\overline{\beta}) \, W_{S^3}(L;\overline{\alpha},\overline{\beta}) \rangle_{BF_k} \, ,
\eea
defines a surgery invariant of $L$ in $M$ and we have:
\bea
\label{surginvrelation}
\langle W_M(L;A,B) \rangle_{BF_k} = { \Upsilon_k(M;L)  \over \Upsilon_k(M) } \, .
\eea
This relation is totally similar to what happens in the $U(1)$ Chern-Simons context \cite{GT3}.

\section{Conclusion}
We now have a full set of results concerning the $U(1)$ Chern-Simons and BF theories. In \cite{MT1} it was shown that in this abelian framework the property $\tau_k(M) = |RT_k(M)|^2$ only holds true for $k$ odd due to the fact that the category $\mathcal{Z}(\mathbb{C}^{\mathbb{Z}_k})$ is modular only in that case. Yet, our abelian framework provides a nice and simple example of Turaev-Virelizier theorem according to which the Turaev-Viro invariant based on a spherical category $\mathcal{C}$ is equal to the Reshetikhin-Turaev invariant of the Drinfeld center, $\mathcal{Z}(\mathcal{C})$, of $\mathcal{C}$ \cite{TuVi}. It has to be pointed out that although $\mathcal{Z}(\mathbb{C}^{\mathbb{Z}_{4k}})$ is not modular a Reshetikhin-Turaev-like invariant can be constructed \cite{Tu,MOO,MT1} and that it coincides, up to some normalization, with the $U(1)$ Chern-Simons partition function. Of course this invariant is not the $\mathbb{Z}_k$ Turaev-Viro invariant. Conversely although the Turaev-Viro invariant based on $\mathcal{Z}(\mathbb{C}^{\mathbb{Z}_k})$ identifies, up to a normalization, with the $U(1)$ BF partition function on the one hand, and with the Reshetikhin-Turaev invariant based on $\mathcal{Z}(\mathbb{C}^{\mathbb{Z}_{k}})$ on the other hand, there is a priori no Chern-Simons theory whose partition function coincides with this last invariant. The only Quantum Field Theory which is related to this Reshetikhin-Turaev invariant is precisely the $U(1)$ BF theory.

Let us end by noticing that although we have identify surgery formulas in the abelian context of the BF theory, to our knowledge such surgery formulas have never been written in the non-abelian (ex. $SU(2)$) context.

\vfill\eject


\begin{thebibliography}{MT03}

\bibitem{W1}
Witten E., Quantum f\/ield theory and the Jones polynomial,
{\it Comm. Math. Phys.} {\bf 121} (1989), 351--399.

\bibitem{GT1}
E. Guadagnini and F. Thuillier, {\it Deligne-Beilinson Cohomology and Abelian Link Invariants}, SIGMA {\bf 4}, 078 (2008).

\bibitem{GT2}
E. Guadagnini and F. Thuillier, {\it Three-manifold invariant from functional integration}, J. Math. Phys. {\bf 54}, 082302 (2013).

\bibitem{GT3}
E. Guadagnini and F. Thuillier, {\it Path-integral invariants in abelian Chern-Simons theory}, Nucl. Phys. B {\bf 882}, 450--484 (2014).

\bibitem{MT1}
P. Mathieu and F. Thuillier, {\it Abelian BF theory and Turaev-Viro invariant}, J. Math. Phy. {\bf 57}, 022306 (2016); doi: 10.1063/1.4942046.

\bibitem{MT2}
P. Mathieu and F. Thuillier, {\it A reciprocity formula from abelian BF and Turaev-Viro theories}, published in "Eulogy for Raymond", Nucl. Phys. B {\bf 912},  327--353 (2016).

\bibitem{RT}
N. Y. Reshetikhin and V. G. Turaev, {\it Invariants of 3-manifolds via link polynomials and quantum groups}, Invent. Math. {\bf 103}, 547--597 (1991).

\bibitem{TV}
V. G. Turaev, O. Yu. Viro, {\it State Sum Invariants of 3-Manifolds and Quantum 6j-symbols},
Topology {\bf 31}, 865--902 (1992).

\bibitem{TuVi}
V. Turaev and A. Virelizier {\it On two approaches to 3-dimensional TQFTS}, arXiv:1006.3501v5 [math.GT] (2013).

\bibitem{BW}
Barrett, J. and Westbury, B., {\it Invariants of piecewise-linear 3-manifolds}, Trans. Amer. Math. Soc. {\bf 348} (1996), 3997--4022.

\bibitem{BK}
Balsam, B. and Kirillov, A., {\it Turaev-Viro invariants as extended TQFT}, available at arXiv:1004.1533.

\bibitem{Tu}
Turaev, V.G., {\it Quantum Invariants of Knots and 3-Manifolds}, de Gruyter Studies in Mathematics, Vol.~10, Walter de Gruyter \& Co., Berlin, 2010.

\bibitem{MOO}
Murakami, H., Ohtsuki, T. and Okada, M., {\it Invariants of three-manifolds derived from linking matrices and framed links}, Osaka J. Math. {\bf 29} (1992), 545--572.

\bibitem{GMM}
Guadagnini, E., Martellini, M. and Mintchev, M., {\it Wilson lines in Chern--Simons theory and link invariants},
Nuclear Phys. B {\bf 330} (1990), 575--607.





\end{thebibliography}
\end{document}